\documentclass[sigconf]{acmart}
\AtBeginDocument{%
  }

\begin{document}

\acmYear{2024}\copyrightyear{2024}
\acmConference[ACM FAccT '24]{ACM Conference on Fairness, Accountability, and Transparency}{June 3--6, 2024}{Rio de Janeiro, Brazil}
\acmBooktitle{ACM Conference on Fairness, Accountability, and Transparency (ACM FAccT '24), June 3--6, 2024, Rio de Janeiro, Brazil}
\acmDOI{10.1145/3630106.3658951}
\acmISBN{979-8-4007-0450-5/24/06}

\title[How Claim Replicability Can Help Bridge the Responsibility Gap]{From Model Performance to Claim: How a Change of Focus in Machine Learning Replicability Can Help Bridge the Responsibility Gap}

\author{Tianqi Kou}
\email{tfk5237@psu.edu}
\orcid{0009-0003-3204-2352}
\affiliation{%
  \institution{Penn State University}
  \city{University Park}
  \state{PA}
  \country{USA}
}

\renewcommand{\shortauthors}{Tianqi Kou}

\begin{abstract}
   Two goals – improving replicability and accountability of Machine Learning research respectively, have accrued much attention from the AI ethics and the Machine Learning community. Despite sharing the measures of improving transparency, the two goals are discussed in different registers - replicability registers with scientific reasoning whereas accountability registers with ethical reasoning. Given the existing challenge of the responsibility gap – holding Machine Learning scientists accountable for Machine Learning harms due to them being far from sites of application, this paper posits that reconceptualizing replicability can help bridge the gap. Through a shift from \textit{model performance replicability} to \textit{claim replicability}, Machine Learning scientists can be held accountable for producing non-replicable claims that are prone to eliciting harm due to misuse and misinterpretation. In this paper, I make the following contributions. First, I define and distinguish two forms of replicability for ML research that can aid constructive conversations around replicability. Second, I formulate an argument for \textit{claim-replicability's} advantage over \textit{model performance replicability} in justifying assigning accountability to Machine Learning scientists for producing non-replicable claims and show how it enacts a sense of responsibility that is actionable. In addition, I characterize the implementation of claim replicability as more of a social project than a technical one by discussing its competing epistemological principles,  practical implications on \textit{Circulating Reference}, \textit{Interpretative Labor}, and research communication.
\end{abstract}



\received{18 Jan 2024}
\received[accepted]{29 Mar 2024}

\maketitle

\section{Introduction}
In recent years, the AI ethics community has produced much literature on improving Machine Learning (ML) transparency. On the one hand, transparency measures can serve the goal of improving accountability. For this goal, transparency measures focus on making artifacts, actors, and development processes open for external auditing and regulations to make prevention, identification, and mitigation of harms easier, and to hold relevant parties accountable. On the other hand, transparency measures have been called for to uphold replicability – maintaining the scientific rigor and integrity of ML research \cite{ashurst2022disentangling} by ensuring that the research process and artifacts are adequately shared to facilitate re-run of studies for verifying the study’s finding’s validity \cite{sep-scientific-reproducibility}.

\subsection{Transparency Measures and the Origins of Concern}
The narrative of \emph{transparency for accountability} initiated from public concerns over the increasing harms generated from ML research and technological systems, which garnered attention from domain experts, policymakers, civil society, AI ethics scholars, and ML scientists from academia and industry. Discerning clear lines of responsibility for harms generated from a complex system that involves numerous parties has been a challenging undertaking, and this obstacle has been famously named the responsibility gap \cite{Goetze2022} - who is responsible for this harm and what does the responsibility call for? As such, developing a mature accountability mechanism that can address wide-ranging harms and satisfy diverse stakeholders is still in progress, despite advances made in computer science, social sciences, and other AI Ethics contributing fields.

The narrative of improving transparency to improve replicability gained traction after the identification of the replication crisis – many fields have observed discrepancies between the results from original study and its replication study \cite{hope2021there, klein2018many, dreber2019statistical}. These concerns exist in the sciences \cite{hope2021there, klein2018many, dreber2019statistical, leonelli2018rethinking, leonelli2023philosophy, national2019reproducibility} as well as in ML  \cite{liu2020replication, pineau2021improving, kapoor2022leakage, raji2021ai, warden2018machine}. In response to these concerns, the Open Science Movement was initiated by leading institutions such as DARPA and ACM; top ML conferences such as KDD, NeurIPS, ICLR, and ICML, etc., have established submission and review guidelines to improve transparency about research artifacts and assumptions.

Taken together, one might argue that transparency can be mobilized toward both goals. Existiting transparency measures include - transparency of decision process \cite{Kroll2021,GDPR}, data development/documentation/curation \cite{Diaz2022, Poirier2022, Papakyriakopoulos2023,Pushkarna2022,Hutchinson2021,Stoyanovich2020,Holland2018}, transparency of model statistics, assumptions, failure modes \cite{mitchell2019model, jacobs2021measurement, milli2021optimizing}; transparency of research limitation and impact \cite{kollnig2022goodbye, smith2022real, ashurst2022ai, robertson2022understanding, kaminski2020algorithmic, metcalf2021algorithmic,reisman2018algorithmic}; transparency of research artefacts such as code and data \cite{boettiger2015introduction, stodden2013best, kapoor2023reforms, albertoni2023reproducibility, li2023trustworthy}; transparency of explanation \cite{balagopalan2022road, barocas2020hidden, bhatt2020explainable, fraser2022ai, lima2022conflict, schoeffer2022there, schuff2022human, shang2022not, zhang2020effect}; transparency of social context \cite{selbst2019fairness, green2020algorithmic, Diaz2022}, transparency of workflow \cite{ludascher2016brief, mcphillips2015yesworkflow, li2023trustworthy, albertoni2023reproducibility}.

Although two goals share measures of transparency, they have different lineages. The concept of accountability arose from moral philosophy, registering with moral or ethical reasoning. Research addressing ML accountability predominantly appears in conferences such as AIES and FAccT, venues with a strong focus on ethics, discussed in tangent with responsibility, blame, and other social values such as fairness and inclusivity. In contrast, replicability has been considered the “gold standard” \cite{lebel2011fearing} and “a key ingredient”\cite{schickore2011does} of science and engineering because it helps establish consistency and regularity, constitutes testability or falsifiability \cite{bogen2000two, popper2005logic}. Although practicing replication traces back to at least as early as the 1600s \cite{shapin2011leviathan}, concerted efforts to interrogate the concept of replicability started around the 1980s \cite{collins1975seven, collins1992changing}. The investigations of the meaning \cite{bogen2000two,sep-scientific-reproducibility,schickore2011does,radder1996and, machery2020replication}, operationalization challenges \cite{collins1992changing, guttinger2019new, radder1992experimental, franklin1998avoiding, fletcher2022replication}, the social dimension \cite{collins1975seven, collins1992changing, feest2016experimenters, hensel2020double}, incentives \cite{romero2019the, mulkay1986replication}, limitations \cite{leonelli2018rethinking,guttinger2020limits, leonelli2023philosophy}, and necessity \cite{feest2019replication, hones1990reproducibility} of replicability remain prominent in \textit{History and Philosophy of Science }and \textit{Sociology of Science}. In this lineage, replicability is usually discussed in tangent with other epistemic values such as \emph{certainty and falsifiability} \cite{popper2005logic} or \emph{objectivity}.

\subsection{Contribution}
Despite the abundance of measures to improve transparency, the adoption rate remains low \cite{Kroll2021} and the problem of the responsibility gap remains salient \cite{Goetze2022, ashurst2022disentangling}. To improve the adoption rate of the above measures for the aim of bridging the responsibility gap, one major challenge is motivating scientists to proactively integrate those tools to engage in social reflection. To such end, this paper conceptualizes a new relationship between the task of improving replicability and the task of engaging in social reflection – social reflections as a pre-requisite of upholding \emph{claim replicability}. \textit{Claim replicability }has advantages over the currently adopted \textit{model performance replicability} to make ML scientists a directly accountable party for harms inflicted by non-replicable claims due to misuse and misinterpretation, positioning social considerations as ML scientists' role responsibility, subsequently helps bridge the responsibility gap.

\subsection{Outline}
The rest of this section defines and distinguishes \emph{model performance replicability} and \emph{claim replicability}. Section 2 lays the foundation for the the main argument by showing the significance of both \textit{model performance claims} (claims to generalizability and robustness of model performance) and \textit{social claims} (ML method’s deliverance of functionality, efficiency, social benefits, etc.), and the relationship between the two types of claims' replicability. 3.1 and 3.2 constitute the argument for claim replicability’s ability to hold ML scientists directly accountable for producing non-replicable claims. Remaining of section 3 address potential rebuttals against claim replicability. Section 4 will discuss claim-replicability's practical implications on \textit{Circulating Reference}\cite{latour1999pandora}, \textit{Interpretive Labor}\cite{graeber2012dead}, and research communication.

\subsection{Definitions of Model Performance Replicability and Claim Replicability}

I define model performance replicability and claim replicability for ML research as the following (Note: for the rest of the paper, I use MPR and CR instead of model performance replicability and claim replicability for concise expression.):
\begin{itemize}
    \item \textit{Model performance replicability: getting the same performance in a replication study.}
    \item \textit{Claim replicability: making the same research claim in a replication study.}
\end{itemize}

\subsubsection{Object of Replication and Two Interpretations of “Result”}
Within the discussion of replicability in the ML community, there exist two beliefs. In the first, a study’s transparency is equivalent to its replicability; and the second, a study is of replicability if the same result is obtained in the re-run. Both beliefs are valid, but they adopt different \emph{objects of replication}. \emph{Object of replication} refers to \emph{the thing that is being replicated}. For the first belief, the object of replication is \emph{the process of a study}; under this belief, the re-run only needs to be able to go through the original study’s procedure to determine the replicability of the original study. This is similar to what \cite{radder1996and} calls “material realization” – all materials and steps of execution are transparently shared, and the same operating condition is obtainable so the replication runner can materially mimic with high fidelity to the orignal study. For the second belief, the object of replication is \emph{result} – the result must be replicated in the re-run to claim that the original study is of replicability. This distinction is important because replication of process does not guarantee replication of results – the same process can lead to different results. 

The key definitional distinction between MPR and CR is the interpretation of \textit{results}. Despite the importance of the distinction as I will show throughout this paper, influential institutions are vague in their official definitions of replicability regarding the meaning of “results”. For example, ACM’s definitions \cite{ACM_2020} use “same measurement”, and \cite{national2019reproducibility} uses “consistent results”. In scholars’ discussion of replicability, the dominant (probably the only) interpretation of results is model performance, such as in \cite{Hutchinson2021, kapoor2022leakage, pineau2021improving, raff2019step, gundersen2018state, ganguli2022predictability, septiandri2023weird, d2022underspecification}.

Despite quantitative interpretation of result as model performance, result can also be interpreted qualitatively. In the context of sciences more broadly, \cite{goodman2016does}’s notion of inferential replicability interprets results as \textit{qualitative conclusions} ,  \cite{gundersen2021fundamental}’s notion of \textit{interpretation replicability}. CR as I defined above is of the same qualitative nature as the other two notions. The different choice of words is a matter of preference to reduce unnecessary confusion because as I will state later a paper usually makes multiple claims, and it would be confusing to say that a paper usually draws multiple \textit{conclusions} or makes multiple \textit{interpretations}.

\subsubsection{Two Characteristics of Claim Replicability}
Besides adopting different interpretations of “result”, two other characteristics of CR set it apart from model performance replicability.

\textbf{Corresponding to individual claim}. MPR is assigned to a whole study because a study usually focuses on a single ML method/model. In contrast, CR correspondes to individual claim because as I will show in sections 2 a single study usually makes multiple claims, and each claim can be of different status in terms of its CR.

\text{\bf A qualitative property}. MPR is commonly expressed as either binary (replicable or not replicable which corresponds to “1/0”) or probabilistic (such as the application of \textit{prediction markets} to evaluate study replicability \cite{liu2020replication}). These expressions are appealing because similar to physical properties such as \emph{melting point} or \emph{density} in physical sciences, they are neat, portable, and easy for comparison. However, this quantitative interpretation has been pointed out to be “too coarse-grained to support replication’s function of evidence amalgamation” \cite{fletcher2022replication}.  \cite{leonelli2018rethinking}'s analysis is another poignant case against the quantitative conceptualization of replicability - "direct replicability" which "is associated with experimental research methods that yield numerical outcomes". Quantitative conception of replicability assumes (or requires) researchers (to) have a high level of control over the materials and procedure through standardization, which incentivizes scientists to focus on reporting standardized procedures while leaving out the idiosyncracies of their study. Contrast to MPR's focus on model performance, CR focuses on evaluating \textit{what is being said} about the model performance. Reliable statements can and should be made even when model performance is not stable.

CR's qualitative nature is derived from the complexity of justification beyond quantitative evidence. CR therefore does not necessitate the form of replication study that strictly mimics key aspects of the original study because standardization of procedure and reporting is only one of many ways to run replications to evaluate CR. To replicate a claim, the replication study can take the form of a separate study aiming at solving the same problem (triangulation). 

\section{Disentangling replicability of model performance claim and replicability of social claim}
CR’s advantage over MPR in helping bridge the responsibility gap relies on CR’s requirement to replicate a diverse set of claims that are made in ML papers – both claims to model performance and claims to the social. Under MPR, the only claim that receives attention for replication is claims to model performance’s generalizability and robustness. However, as 2.1 will show, claims are made not just to model performance, but also often to broader social contexts (ML methods' ability to deliver efficiency of a decision process, functionality of a system, explanatory power, or social benefits such as fairness, etc.). This section points out  1) the existence of social claims, 2) the underestimated significance of the replicability of social claims, and 3) the relationship between replicability of the two types of claim.

\subsection{Diverse Claims in ML Papers – Model Performance Claim and Social Claim}
In \cite{birhane2022values} where the team conducted document analysis on one hundred ML papers from top conferences, they show that the annotated studies foreground values such as performance, robustness, and generalizability, and the main body of the paper focuses on justifying these properties, appealing to the need of the ML community. In contrast, in papers that mention social needs, the connection between the method and those needs is loose and rarely engaged in the main body of the paper. In addition, few qualifications (eg. mentioning limitations) are offered – only two out of the one hundred papers mentioned negative potential impact; even then, they are “abstract and hypothetical” \cite[p.176]{birhane2022values}. For my paper, I apply the language of “claim” to \cite{birhane2022values}’s finding - ML papers foreground model performance claims (generalizability and robustness), and rarely engage in justifying or qualifying social claims.

Social claims are commonplace in ML papers. For example, \cite{birhane2022values} identified commonly included social aspects such as – efficiency, understanding, novelty, real-world applicability, scalability, easiness to work with, fairness, etc. Turning to more concrete examples, \cite{zhang2023trim} developed a risk level assessment NLP model to classify text messages received from pregnant people. In their paper, in addition to making model performance claims such as “TRIM-AI significantly outperforms state-of-the-art baselines” \cite[p. 2]{zhang2023trim}, they also make social claims to functionality such as “better extract semantic and syntax information from code-mixed sentences as compared to hierarchical neural networks”, “improve their operational efficiency, while lowering the operational costs [of the agents working with the system where their method is embedded]”. In \cite[p. 1]{wang2018eann} where they developed an ML method for detecting fake news, in addition to model performance claims, they also make social claims such as “benefit the detection of fake news on newly arrived events”.

When a claim is "loosely connected" or "rarely engaged" \cite{birhane2022values}, it necessarily means that there is little to no evidence in the paper to uphold the claim, under-investigated, and very likely non-replicable - the \emph{claimed good} cannot be delivered in practice. The widespread low replicability of social claims has led to frustrations and concerns, expressed as calling ML papers for a clearer articulation of how a method can translate to concrete impact in ML papers, such as in \cite{chan2023harms} and in \cite[p. 4]{mccradden2023normative} where they state - "There are many examples of diagnostic aids, tools, and systems that demonstrate strong accuracy but have failed to yield benefits to patients.".

\subsection{The Significance of the Replicability of Social Claim}
The belief that producing replicable model performance is ML scientists' core responsibility is prevalent and the academic reward system demonstrates a strong focus on the production of innovative models with good performance. However, the lopsided attention MPR receives compared to CR does not mean that replicability of social claims is not important. To understand why, we need to look into the impact of social claims.

In discussing using data as evidence for scientific claims, Sabina Leonelli states - "What readers are required to take away from a paper is not the data themselves but rather the empirical interpretation of those data provided by the authors in the form of a claim." \cite{leonelli2009locality}
Scientists communicate with the audience of their work in the form of claims. In a paper that summarizes the process and findings of a study that utilizes statistical experiments, authors cannot communicate their findings with mere numerical information, such as p-value, sample size, confidence level, etc. They must interpret those numbers and form claims in sentences to turn them into \emph{"truth" or usable knowledge}.

ML methods are situated in social contexts \cite{cooper2022making, pinch1984social, jasanoff2004states, latour1983give, latour2013laboratory}. ML scientists’ narrative of “changing the world” \cite{packer2013change} has led to the development of ML algorithms for a variety of social contexts \cite{green2020algorithmic, strickland2019ibm, zimmerman2018teachers, green2021data}. With the growing number of cases of algorithmic harms and functionality failures \cite{raji2022fallacy}, we might ask - would the application of ML methods have occurred if people introduced them knew beforehand that harms or failures would happen, or if people who introduced them did not believe that introducing those methods will improve the state of affair in some way(s)? The answer is usually no (except in rare cases of deception). Introducing ML to sites of application is usually believed to be able to positively help the introducer achieve certain goals, such as introducing a pretrial risk assessment algorithm to reduce bias and workload through automation \cite{vannostrand2009pretrial}.

Where do users of ML methods adopt the belief that they will do good? In our society, “expertise is almost always external: it belongs to someone else and our problem is how to recognize it, access it, and mobilize it.” \cite[p. 46]{shapin2004way} Therefore, the most immediate voices of authority that users of ML methods turn to are the scientists who developed those methods and communicated the significance of their methods in publications. Social claims, compared to model performance claims that are expressed in technical language, become more salient voices of authority that \textit{nudge}, \textit{encourage}, \textit{persuade}, and \textit{enable} the circulation and legitimacy of ML methods into various sites of application.

Consider a hypothetical example. If an ML method produces an accuracy of 90\% and the team of scientists wants their method to be applied for real-world applications or used in future benchmarking comparisons, and they cannot merely drop the number 90\% in the conclusion section. Instead, they must make claims such as “our model’s accuracy of 90\% indicates that our model will outperform existing state-of-the-art methods” or “applying our method is a positive intervention into the societal issue \emph{P} which motivated our project.” Without claims as interpretations of model performance, a method will find it hard to travel outside the laboratory where it was developed to be used in benchmarking or be implemented for tackling practical challenges. Claims (especially social claims) shape the future of the developed method – what sites it travels to, what interpretations can be made, who will use the method, and for what purpose, etc. Therefore, as Heather Douglas stated in \cite[p. 85]{douglas2009science}
\begin{quote}
     [In] scientific work…Making empirical claims should be considered as a kind of action, with often identifiable consequences to be considered, and as a kind of belief formation process.
\end{quote}
In practice, social claims tend to be taken for granted \cite{raji2022fallacy, grill2022constructing}, and poorly justified claims \cite{birhane2022values} have led to abundant failures \cite{raji2022fallacy}. This is particularly relevant given the reality that ML methods are commonly used in a variety of contexts different from their original context of development \cite{ashurst2022disentangling, hancox2020beyond}, the prevalent misuse of ML knowledge \cite{widder2022limits, raji2022fallacy}, and the widely accessible computing power and toolkits.

\subsection{Relationship Between Replicability of Model Performance Claim and Social Claim}
In this subsection, I use two hypothetical examples to demonstrate that model performance replicability does not guarantee claim replicability, dissipating the misconception that MPR implies the validity of the entire study.

\subsubsection{Hypothetical Case I}
A study that developed an ML method to identify misinformation and made the following claims:
\begin{itemize}
    \item Claim 1 (model performance claim): Our model outperforms \textit{state-of-the-art} models in identifying misinformation with an accuracy of 95\%.
    \item Claim 2 (social claim): Our model is the first to significantly reduce the workload of human moderators in identifying misinformation due to its unique feature of interpretability.
\end{itemize}
A replication study is run which yields accuracy of 85\%, instead of the reported 95\%. This discrepancy will invalidate (or at least cast doubt on) \textit{claim 1}. Therefore, the replication runner can say that the study is not of MPR. However, in the replication study, the replication runner found that the deployment of the method decreased the workload of human moderators due to its unique feature of interpretability, although to a lesser degree because more time is needed to compensate for the decreased accuracy from the reported 95\%. The replication runner cannot invalidate \textit{claim 2} (not replicated) because there is no other misinformation detection method with the feature of interpretability. In this example, not choosing MPR over CR will unfairly deny the value of the study.

\subsubsection{Hypothetical Case II}
Take the development of \emph{fair} ML models for decision-making in distributing resources such as student admission in higher education institutions as an example. ML methods in this space sometimes aim to uphold equality through parity across race and/or gender \cite{binns2020apparent}. Consider a hypothetical study that made the following claims:
\begin{itemize}
    \item Claim 1 (model performance claim): Our model delivers high accuracy and predictive parity across racial subgroups.
    \item Claim 2 (social claim): Applying our model will improve the state of fairness.
\end{itemize}
A replication study is run and predictive accuracy parity is obtained in the re-run. Thus, the \textit{claim 1} is replicated. However, this does not mean the same for \textit{claim 2} because the criticism of the algorithmic solutions to fairness being limited and potentially counterproductive \cite{binns2020apparent} stands regardless of the replicability of \textit{claim 1}. In this example, although model performance claim is replicated, social claim is not replicated. Using MPR to evaluate the validity of the study will lead to overestimation of the study’s value and blind users to the harms the method can engender.

\section{How Claim Replicability helps bridge the responsibility gap}
So far, I have clarified the definition of MPR and CR, and pointed out the importance of ensuring replicability of both model performance claims and social claims. The lopsided attention MPR receives demonstrates that under the current \textit{paradigm} \cite{kuhn1997structure} of ML research, it is considered imperative to maintain and refine methods and mechanisms for the evaluation of model performance. In contrast, there exist no analogous standards to evaluate claims. This section will demonstrate how CR can help bridge the responsibility gap by holding ML scientists immediately accountable for their producing non-replicable claims. My argument for holding ML accountable applies \cite{Goetze2022}'s account of moral justification for holding computational professionals accountable for the harms generated by the systems designed by them. 3.3 distinguishes accountability and blame which surfaces tensions between CR and other epistemological perspectives that I argue can be reconciled.

\subsection{Bridging the Gap}
\subsubsection{Vacarious Responsibility and Moral Entanglement}
My account of holding ML scientists accountable for producing non-replicable claims utilizes two concepts from  \cite{Goetze2022, goetze2021moral} – \textit{vicarious responsibility} and \textit{moral entanglement}. Vicarious responsibility “concerns cases where one agent is responsible for the actions or behavior of another agent/entity” \cite[p. 396]{Goetze2022}. For example, parents (the vicarious responsible agents) are vicariously responsible for their children’s misbehavior at a party even though the child has their own agency. Within this relationship between vicariously responsible agents and the entity that they are responsible for, there is a moral entanglement because it is \textit{uncertain} “where one’s own agency ends and where another’s begins.” \cite[p. 397]{Goetze2022}

According to Goetze \cite{goetze2021moral, Goetze2022}, moral entanglement should be conceptualized as a continuum and can be weak or strong \cite{goetze2021moral} - the stronger it is, the stronger the moral obligation for the responsible agent to intervene and take accountability \cite{Goetze2022}. The strength of the moral entanglement depends on how central the aspect of vicariously responsible agents' identity connected with the behavior of the agent or entity they are responsible for is to the vicariously responsible agent. For example, parents' moral entanglement with their misbehaving toddler is stronger than a civilian's moral entanglement with their state that committed wrongs. This is because, according to Goetze, one's role as a parent is often more important than one's identity of being a citizen of a state.

In my argument, I posit that there is a moral entanglement arising from a type of "self-reflexive vicarious responsibility" \cite{goetze2021moral}. This type of vicarious responsibility is implicated in the truck driver case cited in \cite{Goetze2022}, where the truck driver who killed a child in a car accident while driving on a highway, is vicariously responsible for, and morally entangled with, the past version of themself - what could have they done differently to prevent the accident from happening even though they broke no rules, therefore, had no reason to question their actions at the time of the accident? As of now, many non-replicable claims exist in the absence of a mechanism for evaluating claims, therefore only retrospectively being identified as problematic. Claim replicability does not hold a place in the current ML \textit{research repertoire} \footnote{I use research repertoire instead of research paradigm because Kuhn never provided a satisfying definition of paradigm.} \cite{ankeny2016repertoires}:
\begin{quote}
    Well-aligned assemblages of the skills, behaviors, and material, social, and epistemic components that a group may use to practice certain kinds of science, and whose enactment affects the methods and results of research.
\end{quote}

Therefore, in the context of CR, \textit{reflexive vicarious responsibility} is characteristic of the relationship that ML scientists have with their constantly evolving scientific self. The moral entanglement in this various responsibility is strong for two reasons. First, their identity as scientists is central to who they are professionally. Second, engaging in methodological reflection and upholding replicability is a salient professional duty. Acting under such moral obligation is scientists' \emph{role responsibility} \footnote{\textit{Role responsibility} arises when one takes on a particular role in society and thus has additional obligations over and above the general responsibilities we all share.} \cite[p. 72]{douglas2009science}.  Therefore, although non-replicable claims were produced unbeknownst to ML scientists in the past, they nevertheless have a strong moral obligation to intervene, make amends, and better their scientific self. 

\subsection{Taking Responsibility}
The strong moral obligation characterized above begs answers to the question: what do we expect ML scientists, the vicariously responsible agents, to do? This subsection will bring this question home by highlighting the action of claim-making which enacts an actionable sense of responsibility for ML scientists to address the moral burden characterized above.

\subsubsection{Research Claim as Research Product}
We typically think of \textit{ML methods as the research products} of a study, the quality of which can arguably be proxied by MPR. In contrast, CR highlights \emph{research claim as research product}, MPR's myopic focus on model performance does not reach the bar. (There is no definitive measure of claim "quality" as I will show in 3.3, I choose "quality" to suit the metaphor of  "product") Just as ML scientists are responsible for the quality of their method (by attending to generalizability and robustness), they are also responsible for research claims. Viewing research claims as research products brings attention to an under-investigated aspect of ML research - the \textit{molding} of this product or the making of a claim. In the following, I will raise two points of consideration to situate the responsibility in the action of claim-making.

\subsubsection{Imputation of Intentionality}
The first point of consideration is the imputation of “intentionality” \cite{johnson2006computer}.
In \cite{Goetze2022}, Goetze points out computational professionals impute their intentionality into the technological systems  - a technological system being “poised to behave in a certain way” \cite[p. 201]{johnson2006computer} is not accidental \cite[p. 9]{Goetze2022}.

Model performances are of little intentionality because they are inert numerical expressions imbued with an aura of austere objectivity and numerical indifference. In contrast, claims home the intentionality of authors of a paper such as soliciting citation, attracting public and peer attention, or calling for implementation, etc. Authors should be aware of what intentionality they are imputing into a claim and if it is justified by evidence at hand. The consideration of intentionality can help ML scientists make replicable social claims by thinking through how their imputed intentionality determines how their methods might play out in the social world.

\subsubsection{Flexibility in Claim-Making}
The second point of consideration is the flexibility inherent in expressing a claim. Making replicable claims requires taking advantage of such flexibility and showing humility. I will list a few here (and elaborate in 4.3) - excluding claims that lack sufficient evidence, introducing qualifications to claims to sufficiently address the uncertainties around the knowledge claim (for example, by attending to practical challenges and ethical concerns during implementation), and specifying targeted audience of their work to reduce the possibility of misinterpretation and misuse.

Such flexibility in claim-making speaks to several barriers to accountability named by Helen Nissenbaum \cite{nissenbaum1996accountability}. First, the making of a claim is usually at the hands of the authors. When speaking about harms from failure of claim replicability, the \textit{Problem of Many Hands} will not be as thorny. Note that I am not implying tracing harms to a non-replicable claim is always an easy task nor am I implying that all claims are easily made replicable by ML scientists. I am only speaking about responsibility for them to take when they did not utilize tools that already exist within the current \textit{research repertoire}. Second, \textit{Bugs'} \cite{nissenbaum1996accountability} “rhetorical power” \cite[p. 869]{cooper2022accountability} to normalize the existence of errors and unpredictability as unavoidable and acceptable in software, and algorithmic systems, makes it easy to scapegoat ML models when harms arise. ML scientists can wield such \textit{rhetorical power} under MPR - \emph{as great as our method can be, the computational stochasticity (e.g., mathematical randomness, shift of data distribution) still eludes the best of us}. Under CR, however, they cannot (as easily) use such rationale to scapegoat non-replicable claims. 

\subsection{How Challenges in Assigning Blame Surfaces Competing Epistemological Perspectives}
Accountability is different from blame \cite{scanlon2008moral} despite blameworthiness' appeal of strengthening motivation. This subsection details the challenges involved in determining blameworthiness (adopting \cite{scanlon2008moral}'s conception of blame) which indicates tensions between CR and other epistemological perspectives, the reconciliation of which is possible. 

In articulating the meaning of blame, \cite{scanlon2008moral} states that judgments of blameworthiness are made by assessing one’s reasons for holding certain intentions and attitudes that go against the norm of their relationship (“social contract”). Blaming someone is to “respond to this impairment by modifying one’s views on their relationship with the blame” \cite[p. 6]{lima2022conflict} I will center two key components of  \cite{scanlon2008moral}'s  conception of blame - 1) violation of the norm of a relationship (norm of replicability), 2) the intention and reasons for violating the norm (of replicability), and analyze if deviations from the two respectively are unjustified therefore non-acceptable. The overall conclusion is that assertions about blame are context-dependent and have a degree of indeterminacy, which surfaces competing epistemological norms and indicates the actualization of CR requires sociological understandings of science.

\subsubsection{On the Norm of Replicability}
Positioning replicability as a normative ideal appears reasonable since it has been conceived to be foundational to science. However, there exist convincing counterarguments, two salient and connected examples are \cite{dang2021scientific} and one relevant concern from\cite{leonelli2018rethinking}. 

The concern that partially motivated Leonelli's \cite{leonelli2018rethinking} development of alternative conceptions of replicability, is treating replicability as a \textit{normative} ideal to demarcate good and bad sciences, risking the value of non-replicability as starting points of inquiries which is epistemically rewarding. This paper's moralizing around CR does position it as a normative ideal, and sheds a negative light on non-replicable claims. Grounding their analysis in the \textit{norm of assertation}, Dang and Bright \cite{dang2021scientific} show that claims by scientists can be appropriately included in published papers, preprints, presentations, etc, even if "they are false, unjustified, and not believed to be true" \cite{dang2021scientific}. They argue that these claims can lead to epistemic successes and therefore should not be held by the standards of \textit{factive norms, justification norms, and belief norms}. This point is in the same spirit as Leonelli's mentioned above - the normative characterization of CR can refute the fruitfulness of non-replicability.

First, Dang and Bright's target of analysis is what they call \textit{public avowals}, the primary audience of which is other scientists \textit{who have expertise on the subject}. \textit{Public avowal} is contrasted with \textit{public science testimony} - which targets communities within society more broadly. Assigning an ML claim to one of the two categories is context-dependent and in my opinion, is going to be challenging at this stage because we do not have a granular understanding of how diverse communities are interacting with claims in ML papers. Despite such ambiguity, I propose that social claims greatly resemble \textit{public science testimony}. It is rare for the public to have access to a particle accelerator or a laboratory freezer to fulfill their experimental wonders in physics or biology; in contrast, it has become prevalent for ML laypersons (in terms of ML knowledge, they can still be experts in other scientific fields, for example, \cite{kapoor2022leakage}) to self-educate and apply ML for various ends given the much more widely accessible educational and infrastructural resources. For this reason, the decision to grant the leniency that Dang and Bright grant to \textit{public avowals} to claims in ML papers needs to be carefully made.

The implication of these concerns for implementing CR is two-fold. First, we need a more granular understanding of the types of ML claims - distinguishing public avowals and public science testimony, and applying CR as a normative ideal to public science testimony. Currently, ML papers do not put any effort into making such a distinction. Second, in scenarios where assigning one claim to one of the two categories is challenging, a more sophisticated approach should be developed instead of looking away from the harms of non-replicable "public science testimony". The aforementioned ambiguity unfortunately transmits to determining if violating the norm of CR is acceptable because adopting competing norms can be justified in their own right. More explicitly in our context, although authors can prevent harms induced by poorly justified claims and be the directly accountable party, they might still justifiably reject blame and prioritize other epistemological principles to include wrong and unjustified claims - accountable but not deserving of blame.

Zooming out from CR, the dilemma shows that one norm (eg. norm of replicability) hardly captures the full picture of the plethora of norms or perspectives (overlapping and contradicting at times) that govern the complex network of relationships between various communities. The landscape of the norms that ought to govern science and society is evolving and indeterminate, and a running debate in disciplines such as \textit{History and Philosophy of Science}, \textit{Sociology of Science}, and \textit{Science and Technology Studies}, for example, \cite{kitcher2001science, douglas2009science, guston2000between, latour1983give, jasanoff2005technologies, jasanoff2004states, jasanoff2015dreamscapes}. In sections 4.2 and 4.3, I will elaborate on my point on "norm".

\subsubsection{On the Intention of Violating the Norm of Replicability}
Even if we have determined that a paper's violation of replicability is unjustified and unacceptable, discerning the intentions of violating CR is still tricky. Let's start by looking at a simplified and straightforward scenario - authors of an ML paper intentionally make non-replicable social claims to boast their study. This would constitute deception which is unacceptable. In practice, detection of such deception is difficult because there exists no analogous mechanism for evaluating claims and  violation of CR is common practice (see subsection where I discussed research repertoire). It would be unfair to say that ML scientists are deceptive in producing non-replicable claims. Deception aside, another challenge goes back to 3.3.1, scientists might be adhering to a different epistemological ideal therefore the intention can arguably be acceptable.

In sum, 3.3 laid out the uncertainties involved in assigning blame to ML scientists for violating CR. However, I should re-iterate that these certainties do not refute the argument for accountability when harms are induced from non-replicable claims -accountability and blameworthiness are not the same. The responses I offered to the epistemological tensions between CR and the epistemic benefits of non-replicability in the context of ML research, can be reconciled by 1) typification of claims (although challenging) which requires 2) deepening our understanding of how diverse audiences interact with different types of ML research claims, 3) and developing norms of communication. These goals are in no way trivial but will guide us out of the dilemma of having to choose one rewarding principle while forfeiting contradicting ones.

\subsection{Address the Worry Over the Erosion of \textit{the Value-Free Ideal}}
Because CR necessitates consideration of social and ethical values, there is another potential rebuttal to my argument that I must address before I discuss the practical implications of CR which concerns the \textit{Value-Free Ideal}, which has dominated science and engineering since the 1950s \cite{douglas2009science}. Preachers of this ideal make the normative claim that scientists should be preoccupied with upholding \textit{epistemic values} – such as predictive accuracy, explanatory power, scope, and simplicity; and it is beyond scientists’ obligation to consider non-epistemic values such as social and ethical values \cite{levi1960must, levi1962seriousness}. To clarify, ML scientists have been engaging in social reflections in response to scholars’ voices that social values must be incorporated to understand and navigate ML research’s social impacts \cite{sloane2022german, green2020algorithmic, selbst2019fairness, cooper2022making}. ML studies aimed at identifying bias and addressing issues of equality have proliferated, and the overwhelming sentiment is the more, the better.

However, CR’s manner of engaging ML scientists in social reflections is different from existing approaches.  Existing measures of engaging ML scientists in social reflections treat producing good models and engaging in social reflections as different tasks. For example, social reflections appear in the form of \textit{add-ons} - impact statement \cite{ashurst2022ai} (a separate document from the main study) or checklist \cite{kapoor2023reforms, norval2022disclosure} that ML scientists can choose to include in their study, or even more external to a study in the form of code of ethics (e.g. ACM computational code of ethics \cite{ACM_2018}). Scientists are willing to play a part in engaging in social reflections to identify and address ML harms but deem social reflections distinct from their core responsibility of building models. As a result, we observe a sense of “dislocated[-ness]” \cite[p. 1]{widder2023dislocated} of accountability and the inclination to procrastinate \cite{cooper2022making} in ML scientists’ current views of accountability - harms are always “another person’s job, always elsewhere” \cite[p. 1]{widder2023dislocated} which can be “solved later and by others” (\cite{Zittrain_2014} in \cite{cooper2022making}).

In contrast, CR integrates social reflections into \textit{evaluating the validity of knowledge claims} - necessitating social reflections in scientific reasoning and determining the competence of someone as a scientist, rather than merely the competence of ethical reasoning. The worry is that the knowledge produced under the influence of social values is less objective and reliable. To this, Douglas \cite{douglas2009science} argued that introducing non-epistemic values to evaluate the validity of claims \textit{does not} necessarily erode scientific objectivity if the values play an \emph{indirect role}. 

Values playing an \textit{indirect role} means that they should not contradict evidence, rather, they determine the sufficiency of evidence – introducing social values can impose extra requirements of evidence for making claims. Using the example of developing ML algorithms for predicting the chance of recidivism - introducing social considerations will require the social claim of \emph{the algorithm doing good} to incorporate evidence beyond high predictive accuracy or accuracy parity across racial subgroups. Social considerations such as 1) the possible backfiring due to unknown interaction between the algorithm and judges informed by those algorithms and 2) how \textit{good} is interpreted by diverse communities with different ideals and priorities, need to be accounted for to make the claim that \emph{our method does good} replicable. In another example of introducing ML methods into medical diagnosis, social reflections on relationships between ML, patients, and doctors, will reveal that making the social claim that \textit{an ML technology brings better experience in medical settings} requires thinking through how ML technology can render those relationships dysfunctional, such as in the case of \cite{pozzi2023automated} where automatic risk scoring induced \emph{hermeneutical injustice} \cite{fricker2007epistemic} that harms both patients and medical professionals. Introduced in this way, social considerations do not provide “epistemic support” \cite{heil1983believing} in the role of evidence, they instead make the standard of accepting a claim more stringent. Therefore, as social considerations can be introduced into evaluating the validity of knowledge claims without eroding scientific objectivity, rejecting CR on the grounds of eroding the Value-Free Ideal would be unwarranted. 

\section{Claim Replicability's Practical Implication}

\subsection{On Circulating Reference and the Danger of \textit{One-Click Replication}}
In John Downer’s essay titled \emph{When the Chick Hits the Fan: Representativeness and Reproducibility in Technological Tests} \cite{downer2007chick}, Downder emphasizes the tradeoff between representativeness and replicability in technological tests - "T[t]he benefits of replicability ['reproducibility'] come at an epistemic cost, it is impossible to make the test simpler without making it less realistic: we are trading representativeness for replicability ['reproducibility']."

Currently, ML research method is characterized by the \emph{Common Task Framework (CTF)} \cite{donoho201750}  - the standardization of workflow, benchmarking evaluation, infrastructure, and research presentation. "Technological tests” (model performance evaluation) are already unrealistic as they are given the neatly curated nature of published datasets for standard comparison and community-level overfitting \cite{roelofs2019meta}. One byproduct of these standardizations is the pursuit of what the field called \textit{computational replicability} through enforcing \emph{one-click replication} - sharing data and code, as well as the configuration of the environments where code should be run, to ensure the replication runner can speedily generate and compare model performance with \emph{one click}. This inclination is a natural direction if we follow MPR. There is nothing wrong with the desire to make model performance replicate. However, making it the sole imperative will blind us from the current malaise of CTF -  moving the field further on the standardization scale, reducing the utility of ML models \cite{flagg2022reward, wagstaff2012machine} by aggravating the alienation of ML from “the uncertainties and contingencies” \cite{grill2022constructing} that abound in real-world applications.

As a result of trading representativeness for replicability, ML evaluation will be reduced to what Bruno Latour calls \textit{Circulating References} \cite{latour1999pandora} - “standardized measures that can be systematized, compared, and analyzed” \cite[p. 21]{downer2007chick}. Attention is diverted from ML studies' ability to deliver functionality, generate understanding, or improve social welfare, to instead generate performances quickly which serve as yardsticks for future comparison under idealistic and unrealistic conditions. Drawing \cite{mccarthy1997ai}'s Drosophila metaphor used in critiquing the limitations inherent in computer scientists' application of computer chess to understanding human intelligence, \cite{evans2023craft} stated - "It was as if geneticists had focused their research efforts on breeding \textit{Drosophila} to race against each other, he remarked. "We would have some science, but mainly we would have very fast fruit flies" \cite{mccarthy1997ai} (Italic in original). Benchmarking to produce "fast fruitflies" should not be the main goal of ML. CR's requiring social reflections can help bring ML scientists' attention back from the "imagined world" \cite{tinn2023between} to produce "machine learning that matters" \cite{wagstaff2012machine}.

\subsection{On Interpretive Labor}
This section suggests that the action of claim-making be viewed as a form of \textit{Interpretative Labor} \cite{graeber2012dead} which should preferably be taken up by ML scientists instead of being disproportionally relegated to users of their study. The courses of action I recommend in 4.3 should be viewed as tools for ML scientists to perform this labor.

Claim-making as interpretive labor refers to the efforts of interpreting numerical outputs. The distribution of such labor is dynamic. If the interpretation of model performance, for example, what it implies and how it should be used, does not (sufficiently) occur in a paper, then readers or users of the paper will have to form their own interpretation. Note that I am not implying that it is ideal to \textit{obviate} interpretive labor from readers or users of an ML paper, which is neither plausible nor beneficial in practice. For example, part of the interpretive labor involved in introducing a piece of ML technology into classrooms is educators' subject and pedagogical knowledge which is beyond ML scientists' scope of knowledge. Instead, I am referring to the labor within ML scientists' wheelhouse.

For example, when a policymaker applies a statistical study or an ML method that lists only numerical expressions in the paper to justify a policy or belief, they need to build a convincing narrative to the public without the help of experts who produced the numbers. A case in point is the introduction of recidivism prediction algorithm into the judiciary process. Instead of investigating and qualifying the claim to reducing bias and labor through automation, the labor of interpretation fell on people who poorly performed the labor, leading to the perpetuation of racial discrimination. In a hypothetical and unrealistic case, if ML scientists exert substantial efforts into incorporating technical requirements and configurations, ethical concerns, and failure modes to substantiate claims, the labor will be drastically reduced at sites of application.

Applying the concept of \textit{Interpretive Labor} to the analysis of CR,  I argue that there exists \textit{structural violence} between ML scientists and the broader society under the current research paradigm. In \cite{graeber2012dead}, Graeber refers to efforts people need to put into understanding and following bureaucratic rules such as understanding and filling out paperwork as \textit{interpretive labor}. The lopsided distribution of interpretive labor (bureaucractic institutions' employees follow rules straightforward to them while people filling out paperwork fend for themselves in navigating the "stupidity" of those rules), according to Graeber [ibid] is founded on \textit{structural violence} - "forms of pervasive social inequality that are ultimately backed up by the threat of physical harm." The powerless are subjected to physical harm when interpretive labor (such as filling out paperworks) is performed incorrectly - being denied access to basic social welfare. Therefore, "nursing homes or banks" are violent institutions because they are "involved in the allocation of resources within a system of property rights regulated and guaranteed by governments in a system that ultimately rests on the threat of force. "Force," in turn, is just a euphemistic way to refer to violence."

In this paper, the \textit{structural violence} in question is in a more sociology-of-science sense, it refers to the \textit{existing communication norm} between ML scientists and diverse communities in society. \textit{Institution}, therefore, takes on the non-material meaning - "A regulative principle or convention subservient to the needs of an organized community"\cite{institution2023}. The communication norm that undergirds and is reinforced by MPR condones ML scientists' actions of making non-replicable social claims; while leaving the readers or impacted communities two options - 1) taking a leap of faith without performing interpretive labor and risk harming other communities or 2) attempting to interpret the social impact accurately by going beyond their scope of expertise. This \textit{structural violence}, as Graeber rightly pointed out,  makes the powerless sympathize with those in power - \textit{ML scientists' job is producing good performing models and it is unreasonable to demand more from them}; conversely, the sympathy is not reciprocated - ML scientists leave negative downstream effects for stakeholders at the sites of application to navigate, or even worse, to be stomached by passively impacted communities.

One rebuttal readers might raise to comparing the research tradition to institutions such as banks or hospitals is that one has more leeway to abstain from ML but not so much with other infrastructural institutions within our society. I disagree with this. First, the impacted members of civil society most of the time do not have the choice to abstain. Second, given the universalizing character \cite{green2020algorithmic} and the branded epistemic advantage of being "'emptied'\cite{ribes2019learned} of domain affiliation" \cite{slota2020prospecting} that ML is chanted for, it is logical to anticipate ML becoming more dominant in scientific disciplines and civil society - which is manifest ML's funding, speed of growth, and its prominent presence in everyday life.

\subsection{On Research Communication}
\subsubsection{Articulate Claims with Communicative Voice} 
Model performance claims (generalizability and robustness) and social claims (functionality, efficiency, equality, human or societal well-being, etc.) must all be made explicit, and expressed with \textit{intentionality} in mind. Instead of selectively presenting claims in abstract, intro, and conclusion, all claims should be formally presented in one dedicated section. Claims should also be made intelligible to diverse audiences of their work and therefore expressed with a communicative voice. In discussing meaningful measures of transparency, scholars have stressed the importance of knowing the “intended recipient” \cite[p. 679]{norval2022disclosure} of what is being made transparent \cite{corbett2023interrogating, ashurst2022disentangling, yurrita2022towards} which should be legible to “who[-ever] is around the blackbox” \cite{ehsan2021expanding}. Ignoring the diversity of audience will “work[s] to disempower, and ultimately hinders broader transparency aims” \cite[p. 679]{norval2022disclosure}. 

Addressing diverse audience speaks to establishing communicative norms (mentioned in 3.2, 3.3, and 4.2) that can mitigate or eradicate structural violence (4.2) within the current dysfunctional norm. More research on how diverse audiences pick up, interpret, critique, and utilize social and computational claims is needed. Over time, efforts should also be put into understanding what repercussions should occur if ML scientists violate the norm. Specification of repercussions is crucial for establishing \cite{bovens2007analysing}'s conception of \textit{accountability} -" A relationship between an actor and a forum, in which the actor has an obligation to explain and justify his or her conduct, the forum can pose questions and pass judgment, and the actor may face consequences." Currently, what repercussions look like is only clear within the relationship between ML scientists and reviewers.

\subsubsection{Systematize Evidence and Increase Evidential Diversity.}
Toward a mature mechanism for evaluating claim, each claim must be accompanied by a list of supporting evidence \cite{ashurst2022disentangling, smith2022real} and the community should have a standardized list of commonly used evidence (qualitative and quantitative).  \emph{Benchmarking comparison, cross-validation, ablation studies, A/B testing, deployment observation, and field feedback, addressing identified concerns generally and specific to targeted applications, triangulation, and qualifications such as failure modes}. Raising the standard of evidence can help cultivate epistemic humility which is much needed in ML \cite{grill2022constructing}.

\subsubsection{Avoid Open-ended Interpretations.}
In writing, authors should avoid using expressions that tend to elicit open-ended interpretations – words that can bear disparate meanings across domains of applications or communities, such as \textit{work}, \textit{benefit}, \textit{improvement}, \textit{social good}, \textit{intelligence}, etc. For example, the \emph{benefit} that a piece of ML technology brings can be unevenly distributed across communities  \cite{ashurst2022disentangling}  who will therefore form \emph{different interpretations of benefit}. Therefore, if a word choice can lead to questions such as \textit{what do you mean by this?} or \textit{can you elaborate?} in real-life discussions of their work with a particular audience, they should be addressed to a reasonable degree while writing the paper.



\section{Concluding remarks}
This paper defined two notions of replicability - \textit{model performance replicability} and \textit{claim replicability} and argued that prioritizing the latter can help bridge the responsibility gap by enacting a strong professional moral obligation to reduce harms induced by non-replicable (social) claims, and provides actionable suggestions toward developing mature mechanisms of evaluating claims in papers. Suggestions made betray that actualizing CR is more of a sociological project than merely technical in that it requires conceptions of functioning communicative norms that can effectively govern the network of relationships in ML ecosystem and counter the existing structural violence between ML scientists and broad society. Such a project essentially enforces changes in ML \textit{research repertoire} and therefore requires \textit{extensive} efforts which are nevertheless worthwhile because it also facilitates the democratization of ML knowledge production because ML scientists will be obligated to engage in conversations with audiences beyond reviewers of their work.

\bibliographystyle{ACM-Reference-Format}
\bibliography{main}


\end{document}